# Enhancement of superconductivity in organic-inorganic hybrid topological materials

Haoxiong Zhang[1,*], Awabaikeli Rousuli[1,*], Shengchun Shen[1,*], Kenan Zhang[1], Chong Wang[1,4], Laipeng Luo[1], Jizhang Wang[1], Yang Wu[2], Yong Xu[1], Wenhui Duan[1,3], Hong Yao[1,4], Pu Yu[1,3] & Shuyun Zhou[1,3]

[1]State Key Laboratory of Low Dimensional Quantum Physics and Department of Physics, Tsinghua University, Beijing 100084, China
[2]Department of Mechanical Engineering and Tsinghua-Foxconn Nanotechnology Research Center, Tsinghua University, Beijing 100084, China
[3]Frontier Science Center for Quantum Information, Beijing 100084, P. R. China
[4]Institute for Advanced Study, Tsinghua University, Beijing 100084, China

**Inducing or enhancing superconductivity in topological materials is an important route toward topological superconductivity. Reducing the thickness of transition metal dichalcogenides (e.g. WTe$_2$ and MoTe$_2$) has provided an important pathway to engineer superconductivity in topological matters; for instance, emergent superconductivity with $T_c$ ~ 0.82 K was observed in monolayer WTe$_2$ [1, 2] which also hosts intriguing quantum spin Hall effect [3], although the bulk crystal is non-superconducting. However, such monolayer sample is difficult to obtain, unstable in air, and with extremely low $T_c$, which could pose a grand challenge for practical applications. Here we report an experimentally**

**convenient approach to control the interlayer coupling to achieve tailored topological properties, enhanced superconductivity and good sample stability through organic cation intercalation of the Weyl semimetals MoTe$_2$ and WTe$_2$. The as-formed organic-inorganic hybrid crystals are weak topological insulators with enhanced $T_c$ of 7.0 K for intercalated MoTe$_2$ (0.25 K for pristine crystal) and 2.3 K for intercalated WTe$_2$ (2.8 times compared to monolayer WTe$_2$). Such organic-cation intercalation method can be readily applied to many other layered crystals, providing a new pathway for manipulating their electronic, topological and superconducting properties.**

## 1.Introduction

MoTe$_2$ and WTe$_2$ in the $T_d$ phase (Fig. 1a) have attracted extensive research interests for their intriguing topological properties both in the bulk crystals and monolayer films. While their bulk crystals are representative three-dimensional (3D) type-II Weyl semimetals [4–6] (Fig. 1b), the corresponding monolayer films are predicted to be two-dimensional (2D) quantum spin Hall insulators [7], with the topological electronic structure [8], edge conduction [9] and quantum spin Hall effect [3] subsequently realized in monolayer $WTe_2$. Introducing superconductivity into such topological materials provide exciting opportunities for topological superconductivity with potential applications in topological quantum computation [10, 11]. Excitingly, monolayer $WTe_2$ exhibits gate-tunable superconductivity with $T_c$ = 0.82 K as compared to its non-superconducting bulk counterpart [1, 2]. Similarly, enhanced superconductivity has also been reported in monolayer $MoTe_2$ with $T_c$ up to 8 K [12, 13] (compared to bulk crystal with $T_c$ of 0.25 K [14]), suggesting that reduced dimensionality plays an important role in the superconductivity. However, so far, such monolayer samples are difficult to obtain through mechanical exfoliation or thin film growth. Moreover, they are extremely sensitive in air [15] and require sophisticated protection layers [1, 2], making it challenging for practical applications. Here we report a strategy to manipulate the interlayer coupling in both $MoTe_2$ and $WTe_2$ bulk crystals through organic cation intercalation (as shown in Fig. 1a), which forms a new class of organic-inorganic hybrid crystals with tailored topological properties, enhanced superconductivity and greatly improved stability compared to monolayer film.

To intercalate large organic cations into these model systems, we employ an electrochemical reaction method using the ionic liquids [$C_n$MIm]$^+$ [TFSI]$^-$ (1-Alkyl-3-Methylimidazolium-Bis (TriFluoroMethylSulfonyl) Imide) (n = 2, 4, ...) as the reacting agents. This is in sharp contrast to conventional electric double layer transistor (EDLT) using ionic liquids [16, 17], where charge carrier concentration is tuned through interfacial electrostatic effect while electrochemical reaction is intentionally avoided. It is interesting to note that recent studies reveal that the electrolysis of water residual within the ionic liquid could also lead to intercalation of oxygen and hydrogen ions [18], which however has negligible effect on the interlayer spacing [19]. In strong contrast, here we achieve successful intercalation of large organic cation (for example, [$C_2$MIm]$^+$ with chemical formula of [$C_6H_{11}N_2$]$^+$) into these layered crystals with dramatically increased interlayer spacing.

The intercalation of organic cation reduces the interlayer coupling of bulk $MoTe_2$ and $WTe_2$, resulting in a band-structural topology transition from a 3D type-II Weyl semimetal to a 3D weak topological insulator [20,21] (schematic in Fig. 1b) - stacking of weakly-coupled 2D quantum spin Hall insulators (see calculated band structure at different interlayer spacings in Extended Data Fig. S1 and experimental monolayer MoTe$_2$ band structure in [22]). Therefore, this organic-cation-intercalation method identifies an innovative strategy to control the dimensionality of transition metal dichalcogenides, so as to design novel electronic states, e.g. superconductivity and complex band-structure topology.

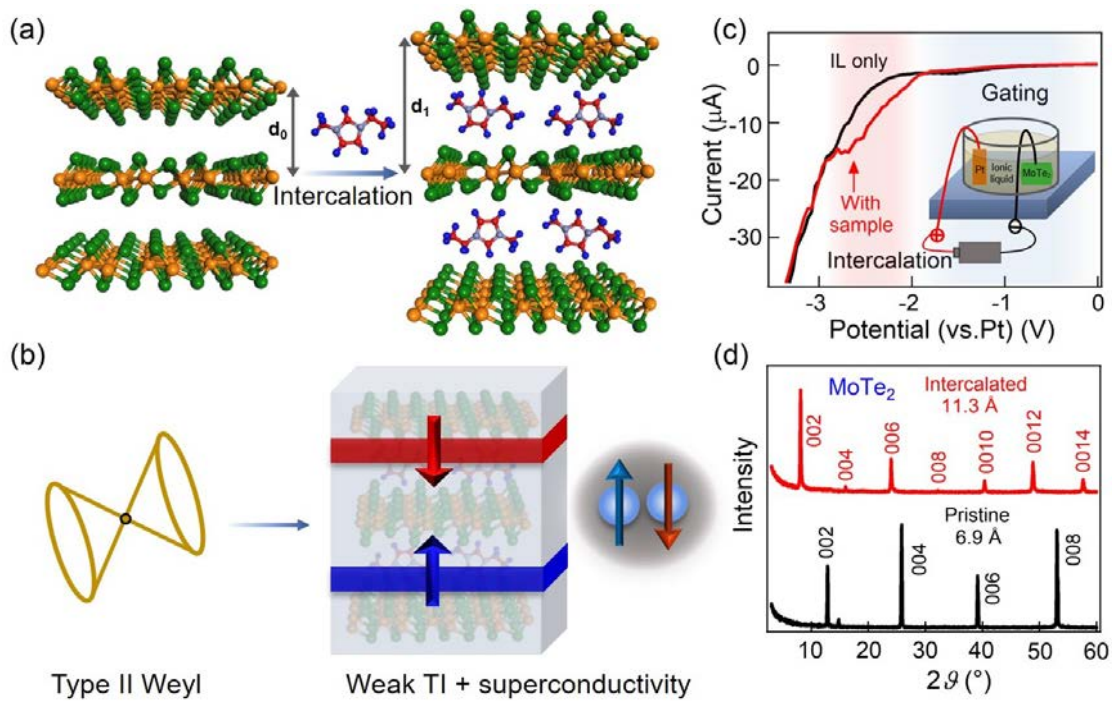

Figure 1: Controlling the dimensionality of transition metal dichalcogenides through organic cation intercalation for tailored topological properties. (a) Schematic illustration for the strategy of controlling the interlayer coupling for the model system of MoTe$_2$ through organic cation intercalation. (b) Schematic for the emergence of superconducting weak topological insulators from type-II Weyl fermion in bulk MoTe$_2$ and WTe$_2$ through ion intercalation. (c) Cyclic voltammetric measurement during the intercalation process. The inset shows a schematic drawing of experimental setup for the electrochemical cell. (d) Direct comparison of XRD results for pristine and an intercalated MoTe$_2$ sample with the interlayer spacing increased from 6.9 Å to 11.3 Å.

## 2. Method

Figure 1c shows a schematic drawing of the intercalation process, in which the single crystal of tens of micrometers thick is immersed into ionic liquid of $[C_nMIm]^+[TFSI]^-$ (see Extended data Fig. S2) and a gate voltage is applied between the sample and the counter electrode (Pt). The clear increase of current at gate voltages between -2 to -3 V (red shaded area of Fig. 1c) during the cyclic voltammetric measurement clearly indicates the induced electrochemical reaction between the sample and the ionic liquid. Using this method, fully intercalated organic-inorganic hybrid crystals with typical thickness of up to 20-100 μm can be obtained within a few hours by taking advantage of two extreme experimental conditions as compared to the conventional ionic liquid gating to facilitate electrochemical reactions. (1) A larger voltage is applied to activate the intercalation process. (2) The environmental temperature is increased to 80-120°C to boost the intercalation efficiency. The successful intercalation is confirmed by X-ray diffraction (XRD) measurements as shown in Fig. 1d, in which a pure phase single crystal was obtained. The clear shift of diffraction peaks to smaller angles indicates a dramatic expansion of the interlayer spacing from 6.9 Å to 11.3 Å (by 64%) for $MoTe_2$ through intercalation. More importantly, these newly formed hybrid compounds are stable (over 25 days without sign of degradation, see Extended Data Fig. S3) in air at room temperature as compared to monolayer samples which require complicated protection layers [1, 2].

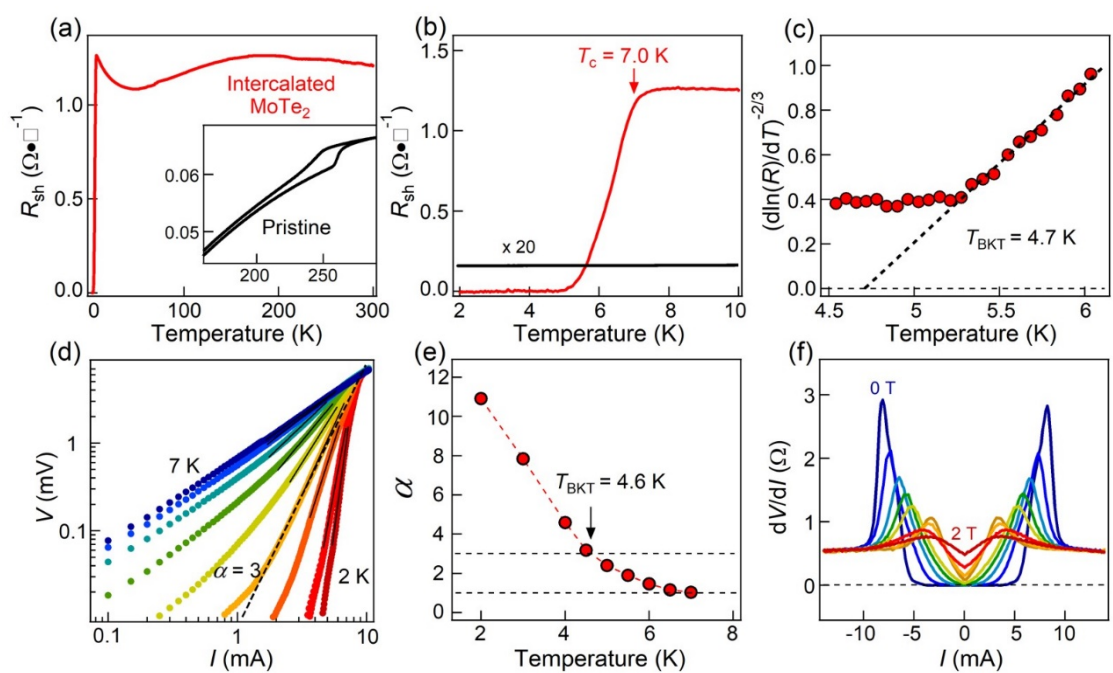

Figure 2: Evidence of emergent superconductivity in intercalated MoTe$_2$ with $T_c$ of 7.0 K. (a) Temperature dependent sheet resistance measurements on intercalated MoTe$_2$ sample. The inset shows zoom-in R-T around the structural transition for pristine MoTe$_2$. (b,c) Low temperature transport measurement of intercalated MoTe$_2$ crystal with the superconducting onset $T_c$ = 7.0 K (b), and corresponding logarithmic plot (c) for extracting $T_{BKT}$ . (d) V-I curves for intercalated MoTe$_2$ at different temperatures from 2 K (red) to 7 K (blue) plotted on a double logarithmic scale. The measurement temperatures are 2, 3, 4, 4.5, 5, 5.5, 6, 6.5 and 7 K respectively. The black lines represent fitted curves with equation of $V \sim I^\alpha$ near the BKT transition temperature. The extended black dashed line (with $\alpha$ = 3) denotes the corresponding $T_{BKT}$. (e) Exponent $\alpha$ obtained from fitting data in (c) near the transition temperature with $V \sim I^\alpha$. $T_{BKT}$

= 4.6 K is obtained at $\alpha$ = 3. (f) Differential resistance d$V$/d$I$ measured at 2 K under different out-of-plane magnetic fields at 0, 0.2, 0.4, 0.6, 0.8, 1.0, 1.2, 1.5 and 2.0 T.

## 3.Results and Discussion

The electrical property of the intercalated MoTe$_2$ sample is revealed by transport measurements. The intercalated sample shows overall larger sheet resistance (Fig. 2a) than the pristine sample (inset in Fig. 2a), suggesting more insulating electronic properties. Meanwhile, the sudden change in resistance at 240-260 K corresponding to the phase transition between $T_d$ and 1$T'$ (different stacking angles) for the pristine sample (inset in Fig. 2a) is absent in the intercalated sample, again indicating that the intercalated sample has weaker interlayer coupling.

Low temperature resistance measurements show that the intercalated MoTe$_2$ samples become superconducting at low temperature. Figure 2b shows an intercalated MoTe$_2$ sample with the highest onset temperature $T_c$ = 7.0 K (defined at the temperature where the resistance reaches 90% of that at the normal state) and zero resistance temperature $T_{c0}$ = 5.1 K. Because of the significantly enlarged interlayer distance in such intercalated system, two-dimensional nature of superconductivity should be expected. Indeed, the linear behavior of $R$-$T$ in Fig. 2c shows that it is consistent with the two-dimensional Berezinskii-Kosterlitz-Thouless (BKT) transition [23, 24] and the extrapolation to $(d\ln(R)=dT)^{-2/3} = 0$ defines $T_{BKT}$ = 4.7 K. Another method to confirm BKT transition is using the $V$-$I$ curves at different temperatures. The $V$-$I$ curves show a power law behavior $V \sim I^{\alpha}$, where $\alpha$ is the slope of $V$-$I$ curves plotted in double logarithmic scale as shown in Fig. 2d. The extracted $\alpha$ in Fig. 2e increases from $\alpha = 1$ at the superconducting onset temperature to $\alpha = 3$ at $T_{BKT}$ = 4.6 K, consistent with the $T_{BKT}$ value extracted from $R$-$T$ curve. Here $T_{BKT}$ is only slightly lower than $T_{c0}$, further supporting that the interlayer coupling in the intercalated sample is very weak. Figure 2f shows the differential resistance $dV/dI$ measured at 2 K under applied magnetic field. The obvious plateau at zero magnetic field gives a critical current of $I_c$ = 6.6 mA, which is more than four orders of magnitude higher than the value ($\approx$ 100 nA) reported in monolayer $WTe_2$ [2]. Such large critical current is in line with the bulk superconductivity in this intercalated sample with thickness of approximately 20 μm compared to a few Å for monolayer film, implying that the intercalated $MoTe_2$ sample contains many superconducting layers that are weakly coupled.

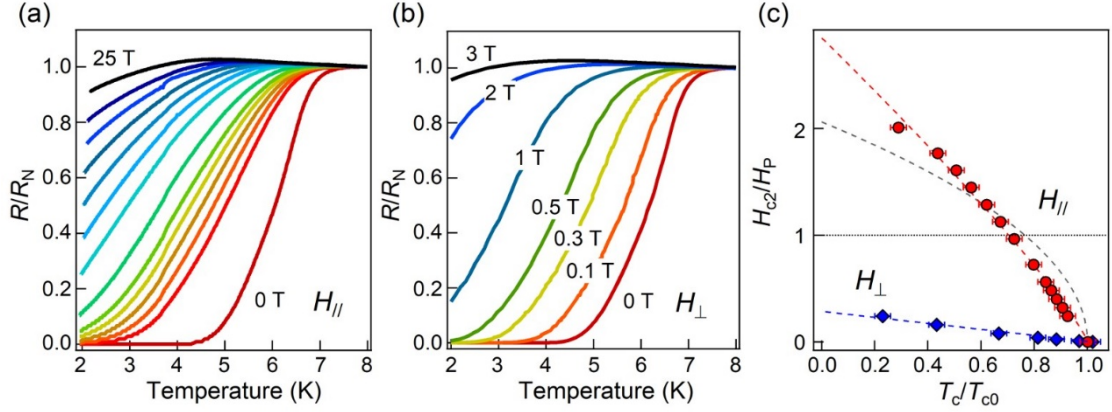

Figure 3: Determination of the upper critical magnetic fields for intercalated superconducting MoTe$_2$. (a,b) Resistance under different in-plane (a) and out-of-plane (b) magnetic fields. The data were normalized by the resistance at 8 K (above $T_c$). The in-plane magnetic fields are 3, 4, 5, 6, 7, 9, 12, 14, 16, 18, 20, 22 and 25 T from red to black curve. (c) Extracted upper critical magnetic fields $H_{c2}$ for $H_{//}$ and $H_\perp$ as a function of temperature. Blue dash curve is fit using 2D Ginzburg-Landau equation. Gray and red dash curves are fitted using

$$H_{c2,//}(T) = \frac{\phi_0 \sqrt{12}}{2\pi \xi_{GL}(0) d_{SC}} (1-\frac{T}{T_c})^{-1/2} \quad \text{and} \quad H_{c2,//}(T) = H^*_{c2,//}(1-\frac{T}{T_c})^{1+x}$$

respectively. $H_P$ is the BCS Pauli paramagnetic limit.

Due to the quasi-two-dimensional nature of this hybrid crystal, the superconducting state is much more robust against in-plane magnetic field ($H_{//}$) than the out-of-plane magnetic field ($H_\perp$). Under an in-plane magnetic field of 25 T, $T_c$ drops by 71% (Fig. 3a). However, the superconductivity is much more sensitive to the out-of-plane magnetic field and $T_c$ already drops by 77% under $H_\perp$ of 3 T (Fig. 3b). Figure 3c shows the extracted upper critical magnetic fields as a function of temperature. By fitting the out-of-plane upper critical field $H_{c2,\perp}$ using the Ginzburg-Landau theory [25] $H_{c2,\perp}(\mathrm{T}) = \frac{\phi_0\sqrt{12}}{2\pi\xi_{ab}(0)^2}(1-\frac{T}{T_c})$, where $\Phi_0$ is the magnetic flux quantum, the in-plane coherence length is extracted to be $\xi_{ab}(0) = 9.6 \pm 0.1\ nm$, which is consistent with pressure induced superconducting MoTe$_2$ [14]. It is interesting to note that for the in-plane upper critical field, the expected behavior for two-dimensional superconductivity $H_{c2,//}(\mathrm{T}) = \frac{\phi_0\sqrt{12}}{2\pi\xi_{GL}(0)d_{SC}}(1-\frac{T}{T_c})^{-1/2}$, where $d_{SC}$ is the effective superconducting thickness along c axis, does not fit the data (gray dashed curve in Fig. 3c) for temperature close to $T_c$. Indeed, for temperature close to $T_c$, the in-plane critical field shows more like linear behavior, reflecting 3D nature of superconductivity because the c-axis coherence length $\xi_c(T)$ (2.5 nm at 6.3 K) is much larger than the interlayer distance. When the temperature is further lowered and $\xi_c(T)$ becomes comparable to the interlayer distance (1.4 nm at 4.9 K), the in-plane critical field exhibits feature of 2D superconductivity. Fitting the data globally using $H_{c2,//}(\mathrm{T}) = H^*_{c2,//}(1-\frac{T}{T_c})^{1+x}$ [14] (red broken curve), we obtain $H^*_{c2,//} = 35.5 \pm 0.9\ T$ and x = -0.14 ± 0.03. Such in-plane upper critical magnetic field $H_{c2,//}$ greatly exceeds the Pauli limit $H_P$ = 9.4 T with $H_P = 1.84 T_{c0}$ [26–28].

There are a few possible scenarios for the large in-plane upper critical field. Firstly, we note that tilted Ising superconductivity [12], which is in analogy to Ising superconductivity in monolayer $NbSe_2$ [26] or gated $MoS_2$ [29, 30], has been proposed for monolayer $MoTe_2$ with spin-orbit splitting induced by the substrate. However, since intercalated bulk crystals have inversion symmetry, the spin-orbit splitting is expected to vanish, making this scenario unlikely. Secondly, the large in-plane upper critical field might be a result of two-bands superconductivity, similar to the case of $MgB_2$ [31] and $LaFeAsO_{0.89}F_{0.11}$ [32] superconductors. Considering that $MoTe_2$ is characterized by nearly compensated electrons and holes [33], this is a possible scenario. A third possible scenario is the Klemm-Luther-Beasley theory [34] which describes stacked two-dimensional superconductors coupled through weak interlayer Josephson tunneling with contributions from both Pauli paramagnetism and spin-orbit scattering. In this scenario, the vortices can fit in the interlayer spacing (or intercalated ions), allowing individual $MoTe_2$ or $WTe_2$ layers to remain superconducting at larger in-plane magnetic field.

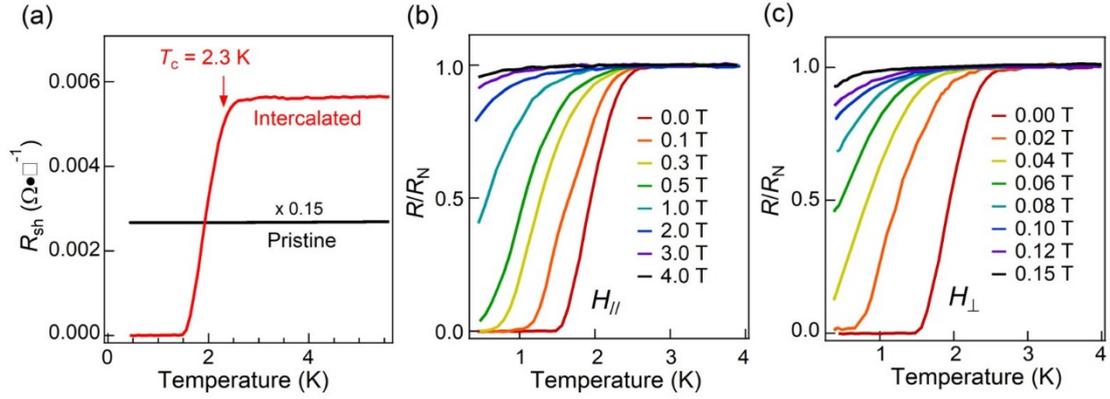

Figure 4: Superconductivity with $T_c$ = 2.3 K in $[C_2MIm]^+$ intercalated $WTe_2$. (a) Direct comparison of temperature dependent resistance measurements on intercalated and pristine samples, showing superconductivity in the intercalated sample with $T_c$ = 2.3 K. (b,c) Temperature dependent resistance measurements under application of in-plane ($H_{//}$) and out-of-plane ($H_\perp$) magnetic fields.

The intercalation method is quite generic and can be extended to other ionic liquids (see Extended Data Fig. S4, S5 for $[C_6MIm]^+$ intercalated $MoTe_2$ with $T_c$ of 6.5 K; Extended Data Fig. S6 for $[TBMP]^+$ intercalated $MoTe_2$ with $T_c$ of 4.7 K), as well as many other layered materials. For example, superconductivity can also be induced in intercalated $WTe_2$, whose pristine counterpart is non-superconducting. The intercalated $WTe_2$ sample shows superconductivity with onset temperature $T_c$ = 2.3 K and reaches zero resistance at $T_{c0}$ = 1.4 K (Fig. 4a). We note that the superconducting onset temperature of the 3D intercalated $WTe_2$ is almost 3 times as compared to monolayer $WTe_2$ ($T_c$ = 0.82 K) [2] which hosts 2D quantum spin Hall effect. This suggests that, in addition to charge carrier doping, intercalation may introduce other effects such as additional phonon modes and Coulomb screening from intercalated organic cations between neighboring quantum spin Hall WTe$_2$ layers which could play an important role in further enhancing $T_c$ compared with monolayer WTe$_2$. Similar to intercalated MoTe$_2$, the superconductivity in intercalated WTe$_2$ is less susceptible to the in-plane magnetic field (Fig. 4b) than the out-of-plane magnetic field (Fig. 4c). While superconductivity survives up to an in-plane magnetic field of 3 T, an out-of-plane magnetic field of 0.15 T is sufficient to destroy the superconductivity.

## 4. Conclusion

The main mechanism of enhanced superconductivity in both intercalated MoTe$_2$ and WTe$_2$ samples is likely attributed to the enhanced density of states near the Fermi level due to the breaking of the balance between the compensated electron and hole pockets by dimensionality and intercalated ions. Moreover, the introduced organic cation between the layers should introduce additional phonon modes which may further boost the $T_c$ [35]. Especially, although the intercalated bulk WTe$_2$ consists of stacked WTe$_2$ layers with very weak interlayer coupling, it exhibits a much higher $T_c$ than gated monolayer WTe$_2$, suggesting that additional phonon modes from organic cation intercalation may play an important role in enhancing $T_c$ from monolayer systems. Moreover, the organic cation might further screen Coulomb interactions between electrons in WTe$_2$ layers which might collaboratively contribute to the enhancement of $T_c$ in intercalated hybrid systems. Thus, our work suggests that reducing dimensionality by the organic cation intercalation forms an important pathway for enhancing the superconductivity. Such enhanced superconductivity in intercalated WTe$_2$, a 3D weak topological insulator, may provide a promising platform to realize robust Majorana zero-mode along step edges on the surface by applying external in-plane magnetic field [36]. Finally, we would like to remark that such intercalation method is quite generic and the extension of such intercalation method to other materials (e.g. TaS$_2$, TaSe$_2$, NbSe$_2$, graphite, etc.) could lead to a wide range of hybrid materials with intriguing properties such as complex electronic band structure topology and superconductivity with enhanced $T_c$.

**Acknowledgement**

This work is supported by the Basic Science Center Program of NSFC (Grant No. 51788104), National Natural Science Foundation of China (Grant No. 11725418, 21975140), Ministry of Science and Technology of China (Grant No. 2016YFA0301004, 2016YFA0301001 and 2015CB921001), and Beijing Advanced Innovation Center for Future Chip (ICFC).

**Author Contributions**

Shuyun Zhou and Pu Yu conceived the research project and designed the experiments. Haoxiong Zhang, Kenan Zhang, Laipeng Luo and Yang Wu grew the bulk single crystal samples. Haoxiong Zhang and Awabaikeli Rousuli intercalated and characterized the samples. Haoxiong Zhang, Shenchun Shen and Pu Yu performed the transport measurements and analyzed the data. Chong Wang, Jizhang Wang, Yong Xu and Weihui Duan performed the first principle calculations and Hong Yao contributed to some theoretical understanding. Haoxiong Zhang, Pu Yu and Shuyun Zhou wrote the manuscript, and all authors commented on the manuscript.

**Author Information**

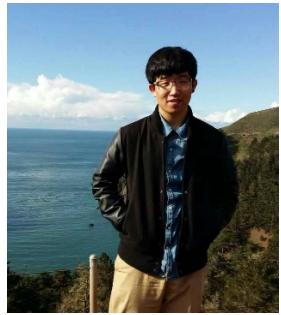

Haoxiong Zhang earned his B.S. degree from Nankai University in 2016. He is now a Ph.D. candidate in the Department of Physics, Tsinghua University. His research focuses on growth and property modulation of quasi-two-dimensional materials and angle resolved photoemission spectroscopy (ARPES) study of their electronic structure, topological property and related quantum phenomena.

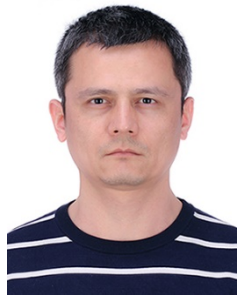

Awabaikeli Rousuli, postdoctoral fellow in department of physics, Tsinghua University, China. He received his PhD degree from Hiroshima University in 2017, Japan. His current research interests focus on growth of two-dimensional materials and investigation of their electronic structures by means of angle-resolved photoemission spectroscopy.

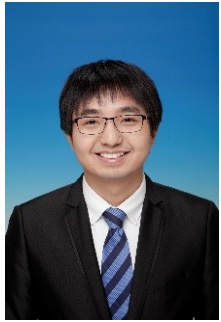

Shengchun Shen, postdoctoral fellow in department of physics, Tsinghua University, received his Ph.D. degree from Beijing Normal University in 2016. His current research interests include magnetism, superconductivity and topological electronics of correlated oxide systems.

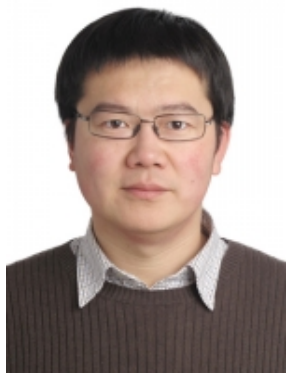

Pu Yu is a professor at department of physics of Tsinghua University. His group focuses on experimental explore of novel phenomena and functionalities in strongly correlated systems, and recently he shows particular research interest on electric-field control of ionic evolution within materials.

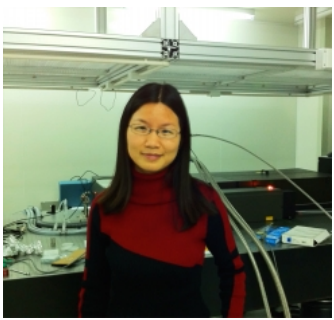

Shuyun Zhou is a professor at the Department of Physics, Tsinghua University. Her research focuses on the electronic structure, property modulation and ultrafast dynamics of two-dimensional materials and heterostructures using advanced spectroscopies, including angle-resolved photoemission spectroscopy (ARPES) and ultrafast time-resolved ARPES.

**Competing Interests**

The authors declare that they have no competing financial interests.

**Equal contributions**

These authors contribute equally to this work.

**Correspondence**

Correspondence and requests for materials should be addressed to Shuyun Zhou (syzhou@mail.tsinghua.edu.cn) and Pu Yu

(yupu@mail.tsinghua.edu.cn).

# Supplement Materials

**Band structure calculation.**

Density functional theory calculations were performed with open MX package [1,2] using the norm-conserving pseudopotentials and the Perdew-Burke-Ernzerhof [3] exchange-correlation functional. The Kohn-Sham wave functions were expanded with a set of pseudo-atomic orbitals (three s orbitals, two p orbitals, two d orbitals and one f orbital for Mo with a 7.0-Bohr cutoff; three s orbitals, three p orbitals, two d orbitals and one f orbital for Te with a 7.0-Bohr cutoff). The Brillouin zone was sampled by a 10 × 10 × 5 mesh.

Figure S1 shows the calculated band structures for $MoTe_2$ with interlayer spacing of 6.9 Å (Fig. S1a, pristine) and 11.3 Å (Fig. S1b, intercalated). Pristine $MoTe_2$ is a type-II Weyl semimetal with four pairs of Weyl points [two independent Weyl points are shown in the inset by red circles with reduced coordinates (0.1037, 0.019, 0.0) and (0.1016, 0.057, 0.0)]. After intercalation, the Weyl points are eliminated completely and band gaps open throughout the Brillouin zone. The dispersion is similar to an inverted quantum spin Hall insulator - monolayer WTe₂. The negligible dispersion along the Γ-Z direction shows that interlayer coupling is weak. In this case, the MoTe₂ can be effectively viewed as stacking of quantum spin Hall insulators, i.e. a weak topological insulator.

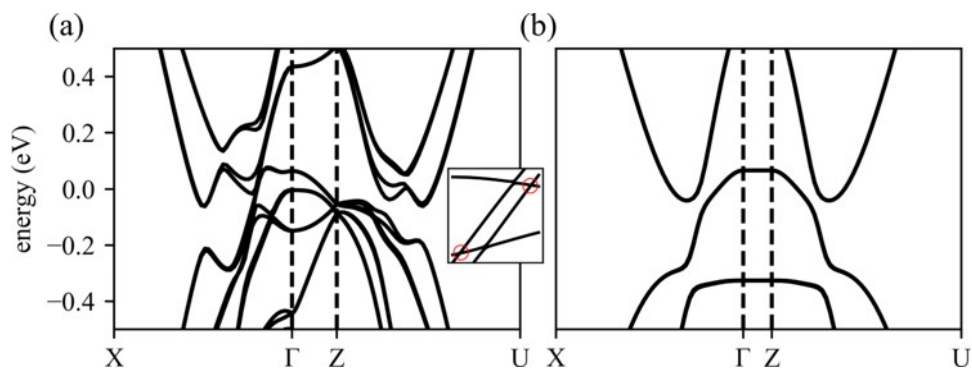

Figure S1: Calculated energy bands for MoTe₂ with different interlayer spacings. (a) Calculated energy bands for MoTe₂ with interlayer spacing d = 6.9 Å. (b) Calculated energy bands for MoTe₂ with interlayer spacing d = 11.3 Å.

## Sample growth and characterization.

High quality pristine 1$T'$-MoTe₂ and $T_d$-WTe₂ single crystals were synthesized by chemical vapor transport method [4]. Polycrystalline 1$T'$-MoTe₂ were first synthesized by heating the high purity stoichiometric mixture of Mo foil (99.95%, Alfa Aesar) and Te ingot (99.99%, Alfa Aesar) at 900°C in a vacuum-sealed silica ampoule for several days. Using chemical vapor transport method with TeCl₄ (5 mg/mL) as transfer agent, polycrystalline 1$T'$-MoTe₂ was transferred from high temperature (930 °C) area to low temperature

(910 °C) area to achieve recrystallization. After keeping the reaction for a few days, the ampoule was quenched immediately in cold water to avoid undesired structural phases. In the end, large and shiny high quality pristine 1$T'$-MoTe$_2$ single crystals were obtained. Upon cooling, the MoTe$_2$ crystal undergoes a structural phase transition from 1$T'$ (stacking angle 93.9°) to $T_d$ phase (stacking angle 90°) at 240 K [4-6]. High quality $T_d$-WTe$_2$ single crystals were synthesized using a similar method.

**Organic cation intercalation.**

Intercalation of ions or atomic molecules has been applied to transition metal dichalcogenides [7] and more recently to ion-based superconductors [8] and black phosphorous [9]. Here using topological semimetals MoTe$_2$ and WTe$_2$ as examples, we demonstrate the effect of organic cation intercalation to induce both superconductivity and band-structure topology change in topological materials. To achieve the organic cation intercalation, the single crystals were placed into a quartz bowl covered entirely with ionic liquid together with a slice of Pt, and then a DC voltage was applied between the sample and grounded Pt (Extended Data Fig. S2). The intercalation starts from the surfaces and edges of the samples, and then the intercalated ions diffuse into the whole single crystal gradually. After the intercalation, the intercalated samples were removed from the ionic liquid immediately and then washed by acetone before further measurements. The intercalated MoTe$_2$ samples show slightly varying interlayer spacings with superconducting onset temperature $T_c$ ranging from 3.7 K to 7.0 K.

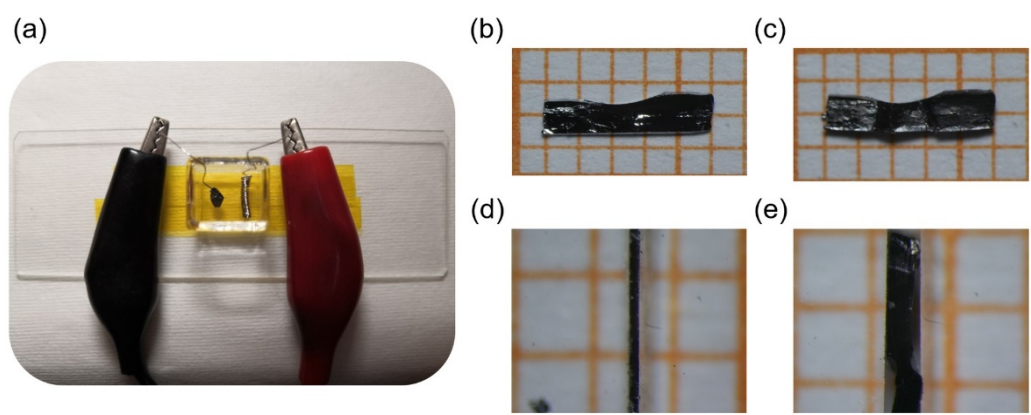

Figure S2: Intercalation process. (a) Picture of the intercalation setup. (b,c) Pictures of the samples before (b) and after (c) intercalation along ab plane of single crystal. (d,e) Pictures of the samples before (d) and after (e) intercalation along c axis of single crystal.

## Transport measurements.

Transport measurements were performed in a Helium-3 Physical Property Measurement System (PPMS, Quantum Design) with the lowest temperature of 400 mK and maximum magnetic field of 9 T. The longitudinal resistances were measured with a four-probe geometry with a small excitation current. The high field magnetic transport measurements were performed in water-cooled magnet with the steady fields up to 33 T in the Chinese High Magnetic Field Laboratory (CHMFL), Hefei.

**Stability of intercalated sample.**

To testify the stability of the intercalated samples, we compare the superconducting properties of as-intercalated MoTe$_2$ sample and sample stored in air for 25 days. The similar $T_c$ between these two measurements suggests that the sample is quite robust in air.

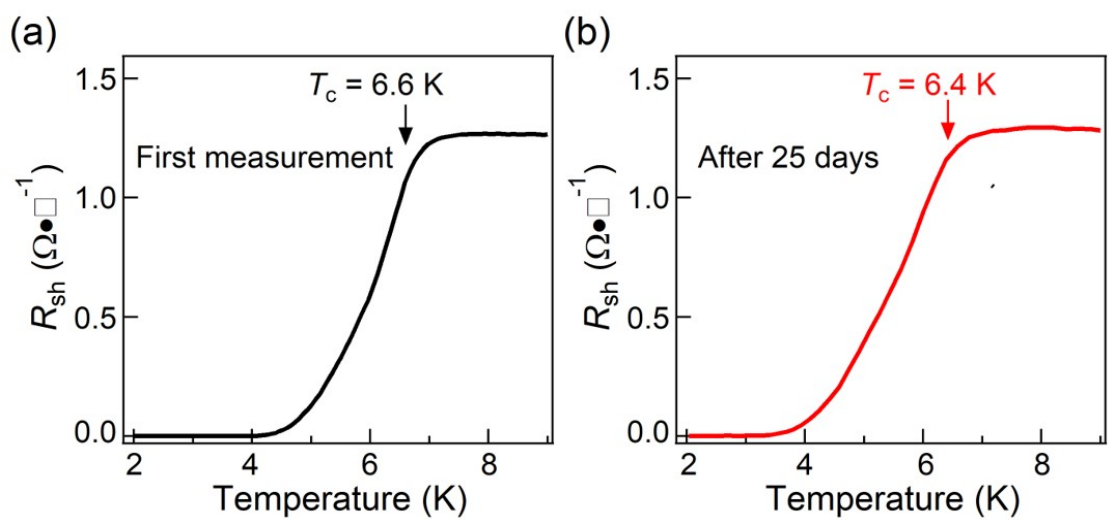

Figure S3: Stability test of [C$_2$MIm]$^+$ intercalated MoTe$_2$. (a) Low temperature transport measurement of intercalated MoTe$_2$ crystal with the superconducting onset $T_c$ = 6.6 K. (b) Low temperature transport measurement of the same sample after 25 days. The superconducting onset is $T_c$ = 6.4 K.

**Intercalation of other ionic liquids.**

Besides [C$_2$MIm]$^+$, the intercalation of other ions is also possible.

For instance, $[C_6MIm]^+$ can also be intercalated into $MoTe_2$ and it shows similar superconducting properties (Extended Data Fig. S4 and S5). The onset superconducting temperature is 6.5 K, and zero resistance occurs at 5 K. Due to the 2D nature, the superconducting state is anisotropic to magnetic field as well. An out-of-plane magnetic field destroys the superconductivity easier. Under an out-of-plane magnetic field of 2 T, $T_c$ drops by 64.5%; while under an in-plane magnetic field of 9 T, $T_c$ drops only by 18.5%.

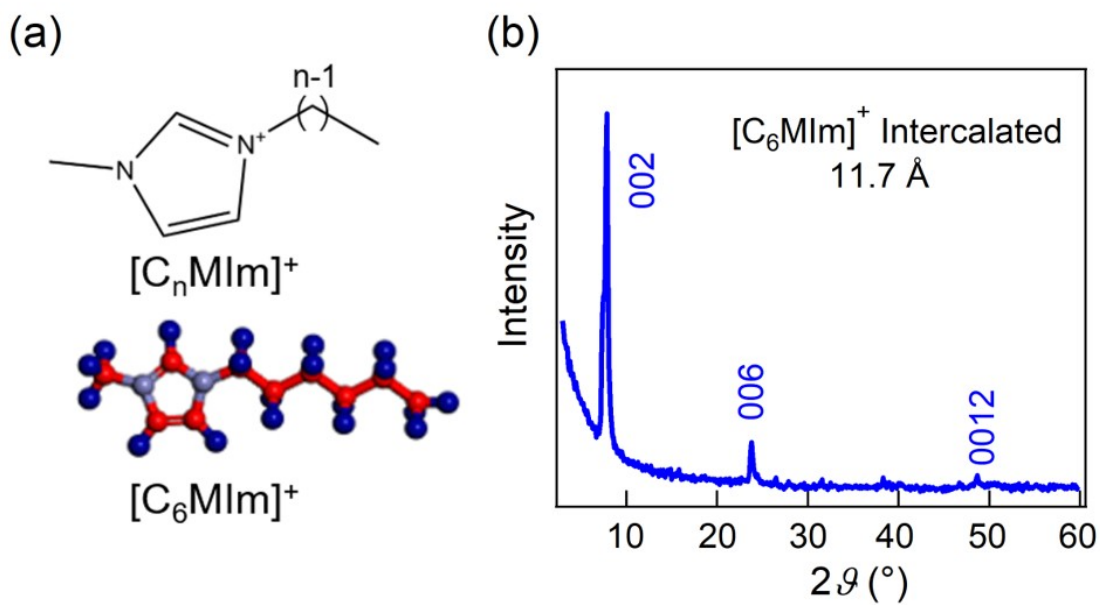

Figure S4: Evidence of $[C_6MIm]^+$ intercalation for $MoTe_2$. (a) Schematic drawing for $[C_nMIm]^+$ (1-Alkyl-3-Methylimidazolium) ionic liquid cations, where n represents the length of the carbon chains. $[C_6MIm]^+$ cation used for $MoTe_2$ intercalation is drew by 3D model. (b) XRD results of $[C_6MIm]^+$ intercalated $MoTe_2$ samples with the interlayer spacing increased from 6.9 Å to 11.7 Å.

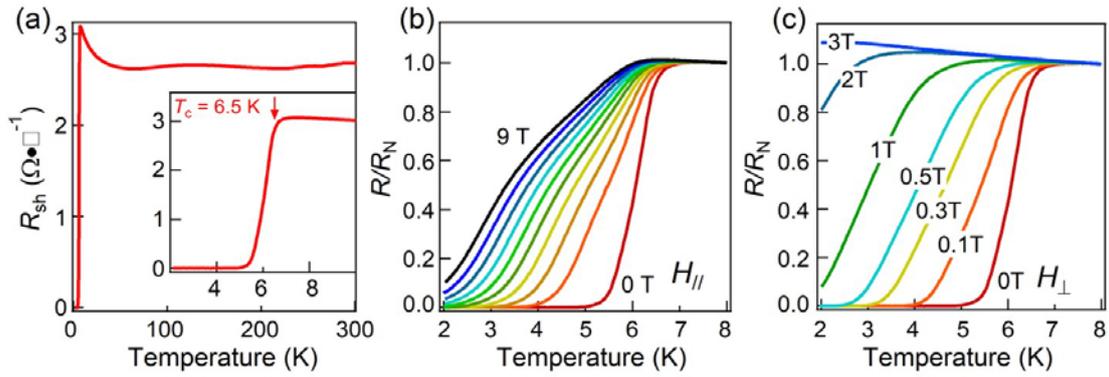

Figure S5: Evidence of emergent superconductivity in $[C_6MIm]^+$ intercalated $MoTe_2$. (a) Temperature dependent sheet resistance measurements on $[C_6MIm]^+$ intercalated $MoTe_2$ sample. The inset shows the superconductivity with onset $T_c$ = 6.5 K. (b,c) Resistance under different in-plane (b) and out-of-plane (c) magnetic fields. The data were normalized by the resistance at 8 K (above $T_c$). The in-plane magnetic fields are 0 (red), 1, 2, 3, 4, 5, 6, 7, 8 and 9 T (black).

The intercalation is not limited to $[C_nMIm]^+$ cations but can also be extended to other families of ionic liquids, for example, phosphonium based $[TBMP]^+$ (Methyltributylphosphonium) cations (see schematic structure in Extended Data Fig. S6b). After $[TBMP]^+$ intercalation, the interlayer spacing increases from 6.9 Å to 20.5 Å (Extended Data Fig. S6a). $[TBMP]^+$ intercalated $MoTe_2$ exhibits superconductivity with onset temperature of 4.7 K (Extended Data Fig. S6c).

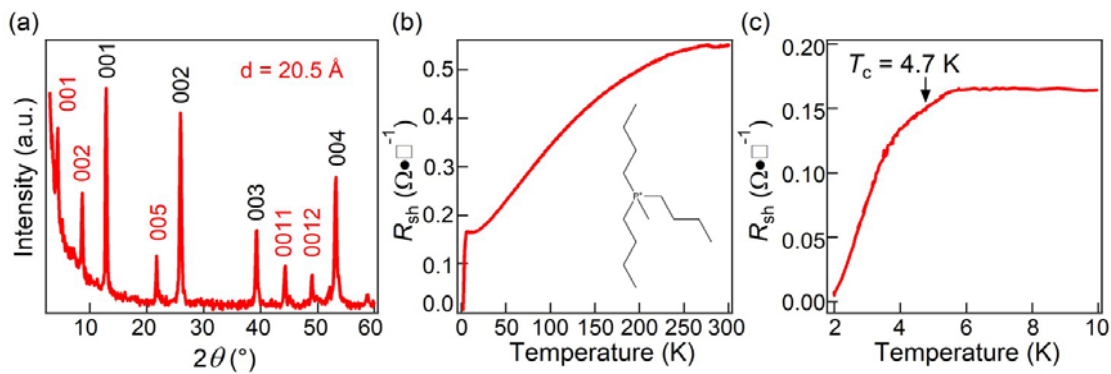

Figure S6: Evidence of [TBMP]$^+$ intercalation for MoTe$_2$. (a) XRD result for [TBMP]$^+$ intercalated MoTe$_2$ with interlayer spacing of 20.5 Å. (b) Temperature dependent sheet resistance measurements on [TBMP]$^+$ intercalated MoTe$_2$ sample. (c) Low temperature transport measurement of [TBMP]$^+$ intercalated MoTe$_2$ crystal with the superconducting onset $T_c$ = 4.7 K.

## Intercalation of WTe$_2$.

Besides MoTe$_2$, intercalation of [C$_2$MIm]$^+$ for WTe$_2$ can also induce superconductivity. According to XRD results in Extended Data Fig. S7, the interlayer spacing increases from 7.1 Å to 10.5 Å after intercalating [C$_2$MIm]$^+$.

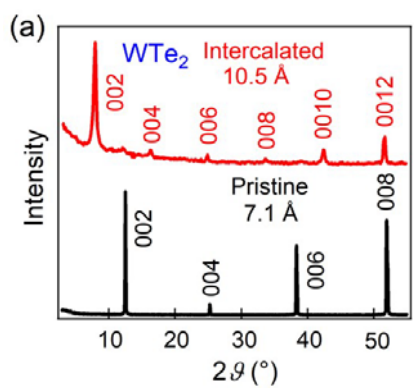

FigureS7: Evidence of [C$_2$MIm]$^+$ intercalation for WTe$_2$. (a) Direct comparison of XRD results for pristine and intercalated WTe$_2$ samples with the interlayer spacing increased from 7.1 Å to 10.5 Å.